\documentclass[journal=jpccck,manuscript=article,linenumber]{achemso}
\usepackage[utf8]{inputenc}
\usepackage[version=3]{mhchem} 
\usepackage{marginnote}
\usepackage{color}
\usepackage{soul}
\usepackage{graphics}
\usepackage{graphicx}
\usepackage{amsmath}
\usepackage{amssymb}
\usepackage{bm}
\usepackage{dcolumn}
\usepackage{longtable}

\author{J. E. P\'{e}rez-Rodr\'{i}guez}
\author{Giuseppe Pirruccio}
\author{Ra\'{u}l Esquivel-Sirvent}
\email{raul@fisica.unam.mx}
\affiliation{Instituto de F\'{i}sica, Universidad Nacional Aut\'{o}noma de M\'{e}xico, Apartado Postal 20-364, M\'{e}xico D.F. 01000, M\'{e}xico}

\title{Spectral thermal  gaps due to plasmon-phonon mode interaction in bilayer systems}

\begin{document}

\begin{abstract}

We present a theoretical study of the modification of the near-field radiative heat transfer due to phonon-plasmon coupling in bilayer systems made of a doped semiconductor and a polar dielectric. By tuning the surface-plasmon mode of the former material in resonance with the phonon modes of the latter one, the near-field spectral radiative heat transfer can be suppressed in a frequency band gap.  We distinguish between the interlayer mode coupling within each bilayer and the intragap one between the two bilayers. We elucidate the role of each of them in the formation of the band gap and we determine the distance range at which one mechanism dominates over the other.  Furthermore, we show that the surface plasmon polariton of the top-most layer allows optimizing the total heat transfer as a function of the separation of the two bilayers.

 \end{abstract}
\maketitle
\date{\today}

\section{Introduction}
Near-field radiative heat transfer (NFRHT) between two bodies is characterized by an increase of the radiated power above the prediction of the Stephan-Boltzmann law. At nanometric distances a dependence of the NFRHT on the dielectric properties of the emitting materials and on the separation between the bodies has been predicted and experimentally observed in plate-plate \cite{Ottens,Kralik1,Kralik2,Guha}, sphere-plate \cite{Sheng12,Shen09} and tip-plate \cite{Kittel08,Kittel13} configuration.

NFRHT is relevant at the micron and submicron scale \cite{Hargreaves,review1,review2}, thus proving crucial in various applications in micro and nanoengineering. The field of thermotronics has been initiated by the introduction of thermal rectifiers \cite{rectifier-grafeno,rectifier-abdallah,Ghanekar:18}, thermal diodes \cite{Otey} and thermal transistors \cite{PhysRevLett.112.044301}. Moreover,  proper engineering of the geometry and materials permits the control of local hot spots and heat sinks \cite{Guha} in nanostructures. On the other hand, from the fundamental point of view, understanding the physics behind the NFRHT has led to the study of the thermal properties of novel materials such as graphene \cite{Liugraphene}, graphene multilayers \cite{PhysRevB.95.245437}, porous materials \cite{Esq17,Thinfilmraul,Biehsporous}, metamaterials \cite{basumeta}. Besides, external magnetic fields have been proposed as a way to actively control the spectral heat flux \cite{PhysRevB.92.125418,spheresh}. The basic idea of these works is to modify the dielectric function of the materials  to change the dispersion relation of light in the cavity or gap between the bodies, thus influencing the NFRHT.

Polariton mode interaction plays a role in NFRHT. For two half-spaces of SiC separated by a gap, the spectral heat transfer is maximum close to the transverse bulk phonon frequency. If one of the half-spaces is replaced by a thin film, the phonon modes split, thus increasing the heat transfer \cite{Francoeur}. By replacing the slabs with layered structures, a super-Planckian behavior arises either from the surface-phonon or hyperbolic modes of the superlattice depending on the dielectric functions of the layers \cite{Guo12}.  If the topmost layer does not support surface modes, the strong enhancement in the NFRHT is not present \cite{Biehs13}.  Metallic films usually exhibit poorer thermal performances in the infrared as compared to dielectric ones, due to their high plasma frequency compared to the thermal one. A way to circumvent this problem is to make use of porosity to redshift the surface plasmon modes \cite{Biehs2007,Esq17,Thinfilmraul}. In this way, NFRHT can be further modified in systems where plasmon and phonon modes coexist in a narrow frequency range. This situation is realized, for example, in bilayers composed of a NaBr substrate coated by a layer of porous Bi. The coupling between the broad gap surface plasmon and the narrow surface phonon mode gives rise to thermal Fano resonances, at which the spectral heat transfer could be suppressed \cite{Esquivel17},  paving the way for further exploration of the control of the NFRHT by engineering the polariton mode coupling at the interfaces of the system.

Thin layers of porous materials may be challenging to fabricate. In contrast, the growth of semiconductor layers such as GaAs, InP, InSb, InAs, AlAs, GaP, Si, or Ge is a well-established technology. By continuously varying their free electron density,  doped semiconductors can be made to behave as dielectrics or as metals. This is achieved in several ways, i.e., by thermal tuning of the intrinsic carrier concentration, ion implantation and optical excitation \cite{Allen,Georgiou}. Due to their naturally low plasma frequency, doped semiconductors are ideal materials to control the NFRHT in the near to far infrared region \cite{dopedsiheat}. For some of these materials plasmon and phonon modes coexist in a narrow frequency range creating a rich system in which the NFRHT can be modified by the formation of new coupled plasmon-phonon modes.

In this work, we show the formation and control of a thermal band gap in the spectral heat flux between two semiconductor-dielectric bilayers separated by a gap. Correspondingly, almost complete suppression of the heat transfer is observed in a large, frequency-tailorable bandwidth. We distinguish between the strong interlayer coupling taking place at each semiconductor-dielectric interface and the intragap coupling between the two bilayers across the vacuum that separates them. We elucidate that the former plasmon-phonon coupling is responsible for the opening of the thermal band gap while the latter provides a common channel for the energy flow. Interestingly, a non-trivial mode competition between these two arises when the gap distance is varied. A mismatch between the resonant frequencies of the two bilayers results in the inhibition of the heat flow, even though each of them sustains a resonant surface mode.

\section{System and materials}

Tuning of the NFRHT relies on a judicious choice of materials that interact in such a way as to enhance or suppress the heat transfer depending on their dielectric functions an geometry. We consider bilayer systems made by stacking a doped-semiconductor (GaAs) and a polar dielectric (NaBr) so that the plasma frequency of the former material lies close to the longitudinal and transverse phonons of the latter one. The thicknesses of each layer are $d_1=50$ nm and $d_2=500$ nm, respectively. We study the NFRHT between two of these bilayers kept at temperature $T_1$ and $T_2$, respectively, and separated by a gap $L$. The system is shown in Fig. 1 from top to bottom, where the left column depicts the individual layers and the total system.

The dielectric functions for the doped GaAs is
\begin{equation}
\epsilon_1(\omega)=\epsilon_{1\infty}-\frac{\omega_p^2}{\omega^2+i\gamma \omega}+\epsilon_{1\infty}\frac{\omega_{1LO}^2-\omega_{1TO}^2}{\omega_{1TO}^2-\omega^2+i\gamma_1 \omega},
\end{equation}
where, $\epsilon_{1\infty}=12.48$ and the first Drude term describes the metallic behavior of the semiconductor with carrier effective mass $m_{eff}=0.064m_e$, plasma frequency $\omega_p=123.1$ THz, and damping $\gamma=18.84$ THz. The second term contains the transverse and longitudinal phonon frequencies $\omega_{1T0}=8$ THz and $\omega_{1L0}=8.54$ THz, respectively, and the damping $\gamma_1=0.072$ THz.
The dielectric function of NaBr is
\begin{equation}
\epsilon_2(\omega)=\epsilon_{2\infty}\frac{\omega_{2LO}^2-\omega_{2TO}^2}{\omega_{2TO}^2-\omega^2+i\gamma_2 \omega},
\end{equation}
where $\omega_{2LO}=39$ THz, $\omega_{2TO}=25$ THz and $\gamma_2=0.4$ THz correspond to the longitudinal phonon frequency, transverse phonon frequency and damping, respectively. The high frequency value of the dielectric function is $\epsilon_{2\infty}=2.6$.

\section{Near field heat transfer}
The near-field heat transfer is calculated within the framework of Rytov's theory of fluctuation electrodynamics \cite{Vinogradov}. Throughout this work, the wavevector component normal to the interfaces in vacuum is $\kappa=\sqrt{\omega^2/c^2-\beta^2}$, and for the $i^{th}$ material is $\kappa_i=\sqrt{\epsilon_i \omega^2/c^2-\beta^2}$, where $\omega$ is the frequency and $\beta$ is the parallel component of the wave vector to
the interfaces.
  The total heat flux $Q(L,T_1,T_2)$ is
  \begin{equation}
  Q(L,T_1,T_2)=\int_0^{\infty} d\omega S_{\omega}
  \end{equation}
where   $S_{\omega}$ is the spectral heat function given by,

\begin{equation}
S_{\omega}(L,T_1,T_2)=\left[ \Theta(\omega,T_2)-\Theta(\omega,T_1)\right ]\sum_{j=p,s}\int\frac{\beta d\beta}{(2\pi)^2} (\tau_{j}^{prop}+\tau_{j}^{evan}),
\end{equation}
where $\Theta(\omega,T)=\hbar \omega /(exp(\hbar \omega/K_B T)-1)$ is the Planckian function.
The sum is over both possible polarizations $p$ and $s$. $\tau_{\omega}^{prop}$ and $\tau_{\omega}^{evan}$ are the densities of the propagating and evanescent modes, respectively, written in terms of the reflectivities $r_{p,s}$  as
 \begin{equation}
 \tau_{p,s}^{prop}=(1-|r_{p,s}|^2)^2/|1-r_{p,s}^2 e^{2i\kappa L)}|^2,
 \end{equation}
 \begin{equation}
  \tau_{p,s}^{evan}=4 Im(r_{p,s})^2e^{-2 |\kappa|  L}/|1-r_{p,s}^2 e^{-2|\kappa|L}|^2.
   \end{equation}
We define the total transmission  as $\tau=\tau_{p,s}^{evan}+\tau_{p,s}^{prop}$.
It is convenient to introduce the spectral heat transfer coefficient \cite{Song15}, $h_{\omega}(T)$, defined as
\begin{equation}
h_{\omega}(T)=\lim_{\Delta T \to 0}\frac{S_{\omega}(T+\Delta T)-S_{\omega}(T)}{\Delta T},
\end{equation}
where $T$ is a mean temperature that we fix at  $T=300$ $K$.

The reflectivities are calculated using the surface impedance approach that captures the interaction of the modes excited in the GaAs and NaBr layers via the transfer matrices of the layers\cite{Cocoletz}.

 \section{Results}
To understand the thermal behavior of the cavity, we first analyze the individual layers and then the bilayered composite system. Since for the NFRHT the main contribution comes from the $p$-polarized waves, we calculate $\tau_p$ for the evanescent region ($\beta>\omega/c$). The thickness of the layers is within the range of values allowed by current experimental techniques \cite{Moon}. Throughout this work the gap separation remains constant and equal to $L=50$ nm. Figure 1 (a) shows two GaAs slabs of thickness $d_1=50$ nm and carrier density $N=3\times10^{23}$ m$^{-3}$, that yields a plasma frequency of $\omega_p=123.1$ THz together to  the corresponding $\tau_p$ calculated as a function of frequency, $\omega$ and normalized wavevector, $\beta/k_0$, where $k_0=\omega_0/c$ with $\omega_0=100$ THz. Due to the metallic character of doped GaAs this system sustains a plasmonic waveguide mode or gap surface plasmon polariton (GSPP) \cite{nanoscale}. The red dashed lines indicate the frequency at which the dispersion of the GSPP flattens, i.e. $\omega_{GSPP}=\omega_p/\sqrt{\epsilon_{1\infty}+1}\sim 33$ THz and the two bulk optical phonon frequencies of GaAs. Similarly, in Fig. 1 (b) two NaBr layers of thickness $d_2=500$ nm and the corresponding $\tau_p$ are shown. The blue dashed lines correspond to the transverse and longitudinal optical phonon modes for the bulk NaBr.  Note that the GSPP lies in the Reststrahlen region of NaBr, i.e., $\omega_{2TO}<\omega_{GSPP}<\omega_{2LO}$, which provides the base for a complex mode interaction in the composite system.

The consequence of the plasmon-phonon hybridization in the NFRHT is shown in Fig. 1(c), where a clear reduction of the mode density leads to the opening of a thermal band gap around $\omega \sim 200$ THz at which $\omega_{GSPP}\simeq\omega_{SPhP}$. We stress that this kind of band gap is not the consequence of spatial translational invariance and Bloch theorem. The opening of the thermal band gap in our non periodic structure is due to destructive interference between the GSPP and the SPhP. 

 This is a purely electromagnetic process related to the interference between the broadband, dispersive GSPP and the narrowband surface phonon polariton, which leads to the opening of a window of suppressed spectral heat flux. Here, we apply this concept to the thermal properties of a nanostructure, demonstrating that the interference between broad and narrow optical modes leads to the opening of an electromagnetically-induced window of suppressed spectral heat flux.

 \begin{figure}[h]
    \centering%
    \includegraphics[width=0.7\textwidth]{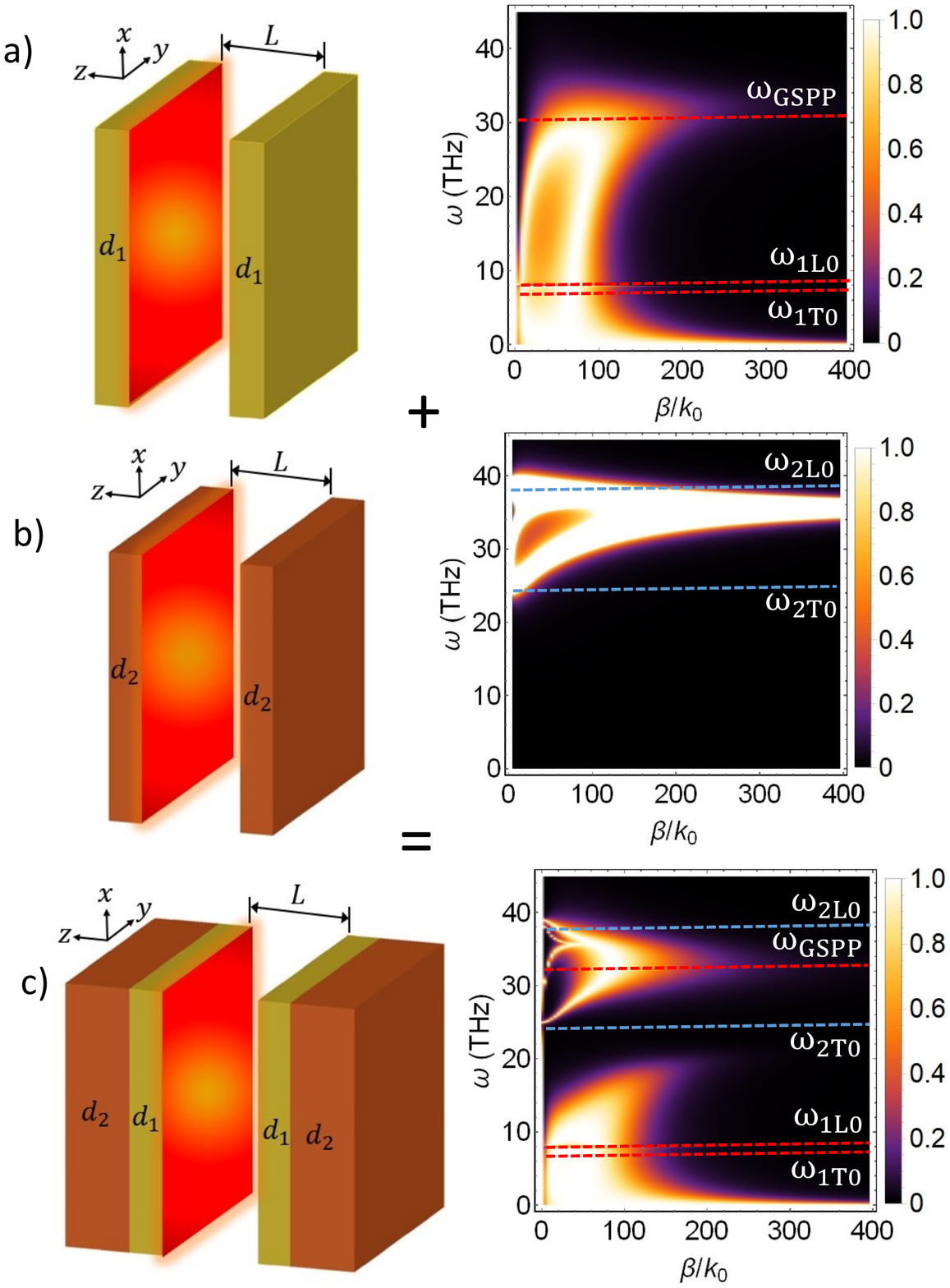}%
    \caption{ (a) Two slabs of doped GaAs with  $N=5\times10^{23}$ m$^{-3}$, of thickness $d_1=50$ nm and plot of $\tau_p$ as a function of the  frequency and normalized wave vector. The red dashed lines corresponds to $\omega_{GSPP}$ and to the frequencies of the longitudinal and transverse optical phonons for two semi-infinite slabs of GaAs. (b) Two slabs of NaBr of thickness $d_2=500$ nm and the corresponding $\tau_p$. The dashed blue lines indicate the position of the transverse and longitudinal optical phonons for two semi-infinite slabs of NaBr. The combined system GaAs-NaBr is shown in  (c). The  transmission coefficient, $\tau_p$,  shows the hybridization of the plasmon mode and the phonon modes of the NaBr. We define $k_0=\omega_0/c$, where $\omega_0=100$ THz.}  \label{fig:pull-in}
\end{figure}

The thermal band gap is further analyzed in Fig. 2 where the spectral heat coefficient $h_{\omega}(T)$ is shown. The value of $h_{\omega}(T)$ is shown for two GaAs layers of thickness $d_1=50$ nm (dashed gray curve), two NaBr layers of thickness $d_2=500$ nm (dotted grey curve) and the bilayer system (continuous black curve). The mean temperature is kept at $T=300$ K. The mode hybridization suppresses the intense response of the GaAs and around $\omega=25$ THz where $h_{\omega}$ shows a minimum value. This behavior occurs since $\omega_{GSPP}$, lies between the longitudinal and transverse phonons. The phase and amplitude condition required to open the gap is realized for frequency values for which the contribution to the NFRHT from the GSPP is much larger than the surface phonon one, leading to the redshift of the thermal band gap with respect to both $\omega_{GSPP}$ and $\omega_{SPhP}$ the optical phonon frequencies.

  \begin{figure}[h]
    \centering%
    \includegraphics[width=0.7\textwidth]{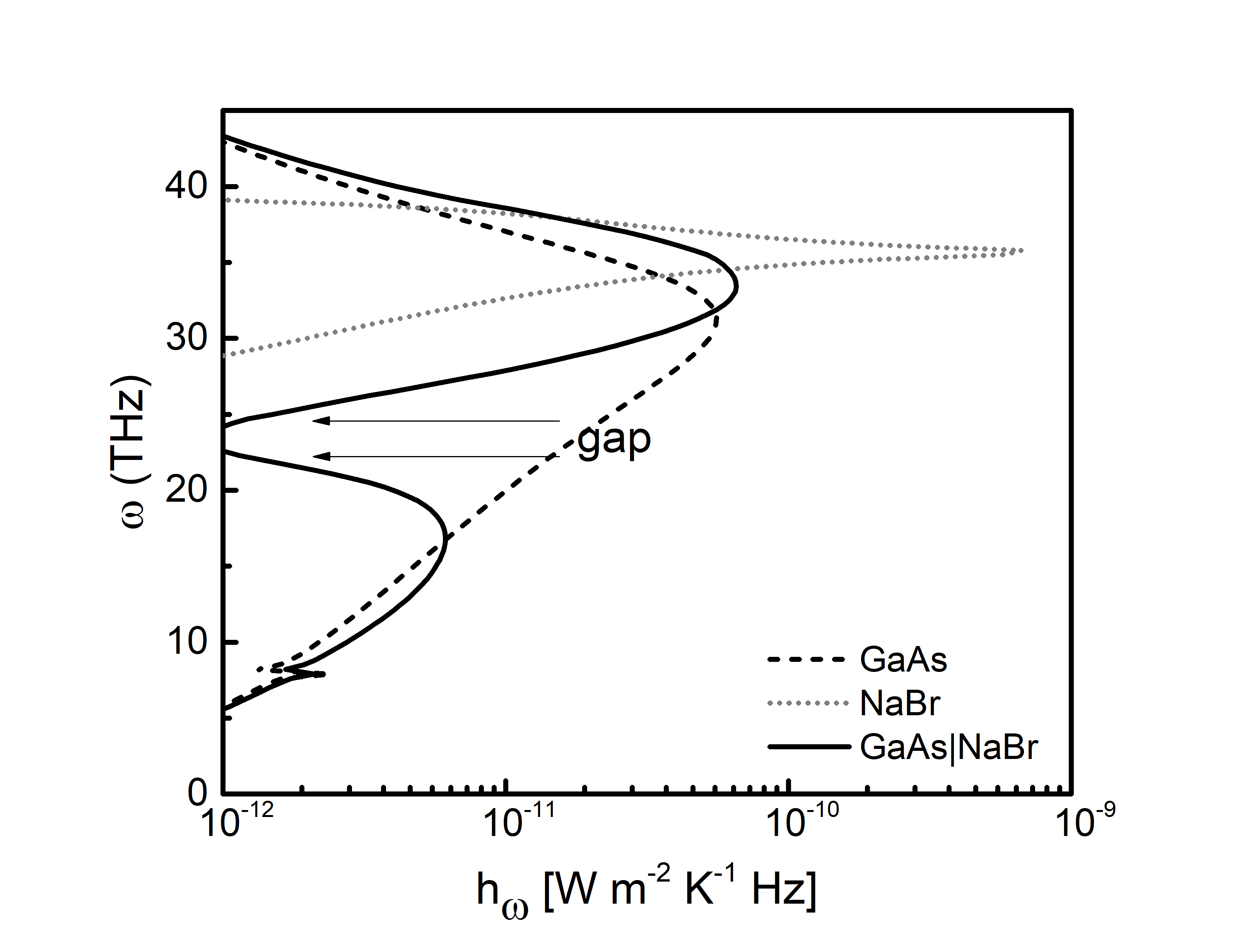}%
    \caption{Spectral heat coefficient $h_{\omega}$ as a function of frequency. The three curves correspond to $h_{\omega}$ for two GaAs slabs (dotted grey curve), for two NaBr slabs (dashed gray curve) and for the bilayer system (continuous black curve).  A sharp decrease in the thermal coefficient at a frequency $\omega \sim 25$ THz, indicated by the arrows, is seen as a consequence of the opening of the thermal band gap.} \label{fig:pull-in}
\end{figure}

The origin of the band gap in the spectral heat transfer can be explained qualitatively by analyzing the dispersion relation of the  surface modes  at the GaAs-air and NaBr-air interfaces and the coupled mode sustained by the  GaAs-NaBr surface.  We plot the frequency \cite{PhysRevLett.71.145} $\omega$ $vs$ $k_x=k_{sp}=\frac{\omega}{c}\epsilon(\omega)/\sqrt{\epsilon(\omega)+1}$, where $\epsilon$ is the dielectric function of either material and for the bilayer the dielectric function for p-polarized light is calculated from \cite{ZAYATS2005131} $\epsilon^{-1}=(d_1/(d_1+d_2))\epsilon_1^{-1}+(d_2/(d_1+d_2))\epsilon_2^{-1}$. The typical values of $N$ we consider are:  $N=5\times10^{22}$ m$^{-3}$. In Fig. 3 (a) the dispersion relation is shown for NaBr, GaAs, the bilayer, and  the light cone. At high frequencies the dispersion relation of the bilayer follows  the dispersion of the polar material, while  at short frequencies it follows the one of  the GaAs. A zoom  is shown in Fig. 3(b), where  in an intermediate frequency region  the bilayer behaves as a hybrid material. Here  the modes  in the propagating region are responsible for the gap in the spectral thermal emission.

\begin{figure}[h]
\centering%
\includegraphics[width=0.9\textwidth]{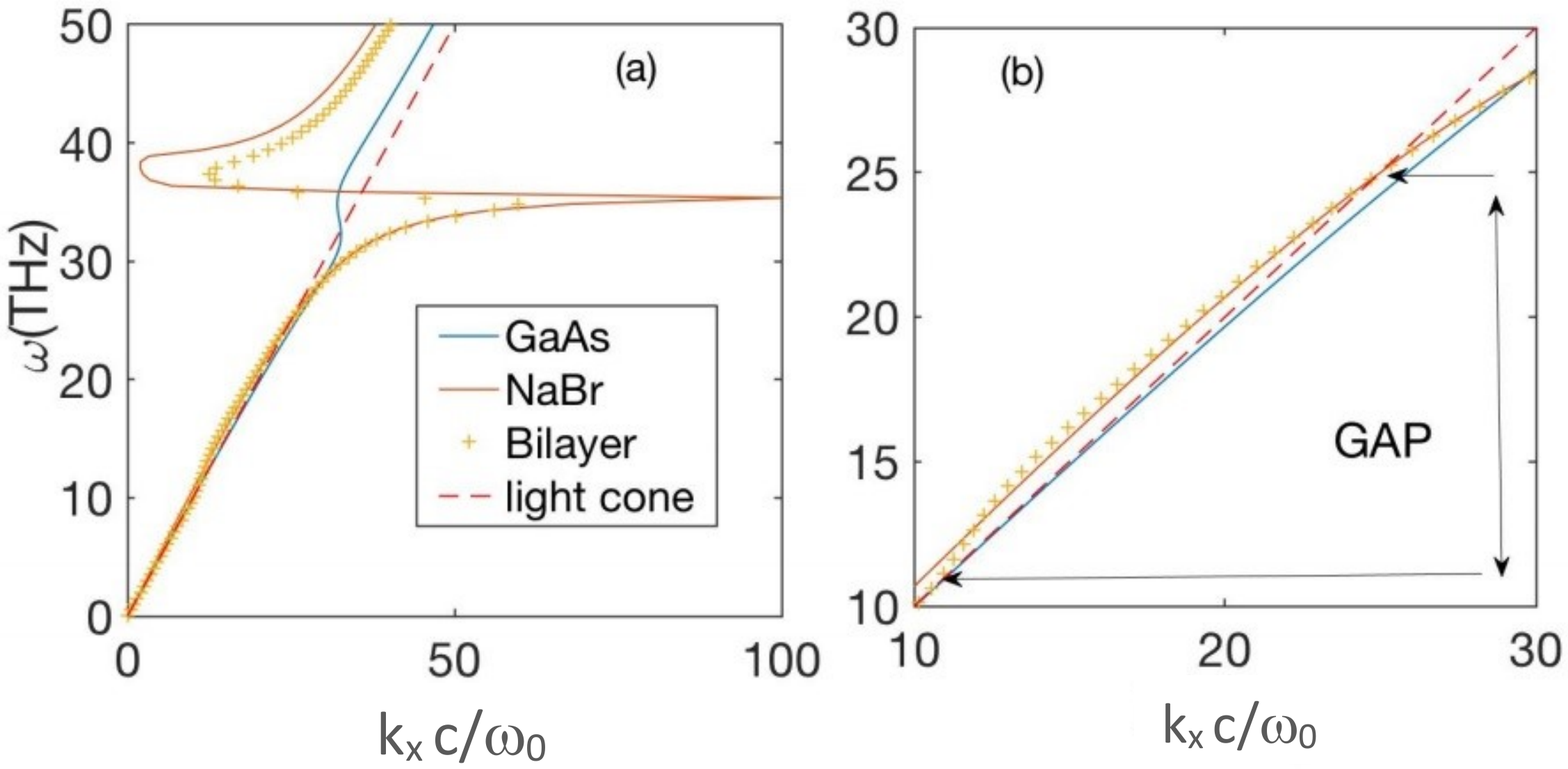}%
\caption{(a) Dispersion relation for a GaAs-air surface (long-dahed curve), a NaBr-air surface (short-dashed curve) and for the GaAs-NaBr surface (crosses). (b) Detail that shows how the  dispersion relation  crosses the light cone into the propagating wave region. The gap where the thermal emission occurs is indicated.} \label{dispersion}
\end{figure}

In Fig. 4 we analyze how the thermal gap forms as the charge carrier density in GaAs changes thus detuning the plasma frequency from the phonon
frequencies of the NaBr, by calculating the  total transmission $\tau$. For the $s$-polarization the dielectric function of the bilayer is given by \cite{ZAYATS2005131} $\epsilon=(\epsilon_1 d_1+\epsilon_2 d_2)/(d_1+d_2)$ (a),  $N=5\times10^{23}$ m$^{-3}$ (b), and $N=5\times10^{24}$ m$^{-3}$ (c), all experimentally accessible doping levels.  For the lowest value of $N$ $\omega_{GSPP}$ lies below $\omega_{TO2}$ of NaBr and, therefore, the thermal gap opens. On the other hand, for the highest value of $N$ $\omega_{GSPP}$ lies well above $\omega_{TO2}$ and no mode coupling occurs. This is confirmed by the spectral heat transfer coefficient shown in Fig. 4(d), where no thermal gap is observed for large carrier densities and the radiative thermal properties of the system are given by the two almost independent contributions from GaAs and NaBr (see Fig. 1). Thus, the doping level provides an efficient and practical way to tailor the thermal band gap and the NFRHT in nanostructures made of doped semiconductors, a characteristic that might be relevant for semiconductor-based devices.

\begin{figure}[h]
\centering%
\includegraphics[width=0.9\textwidth]{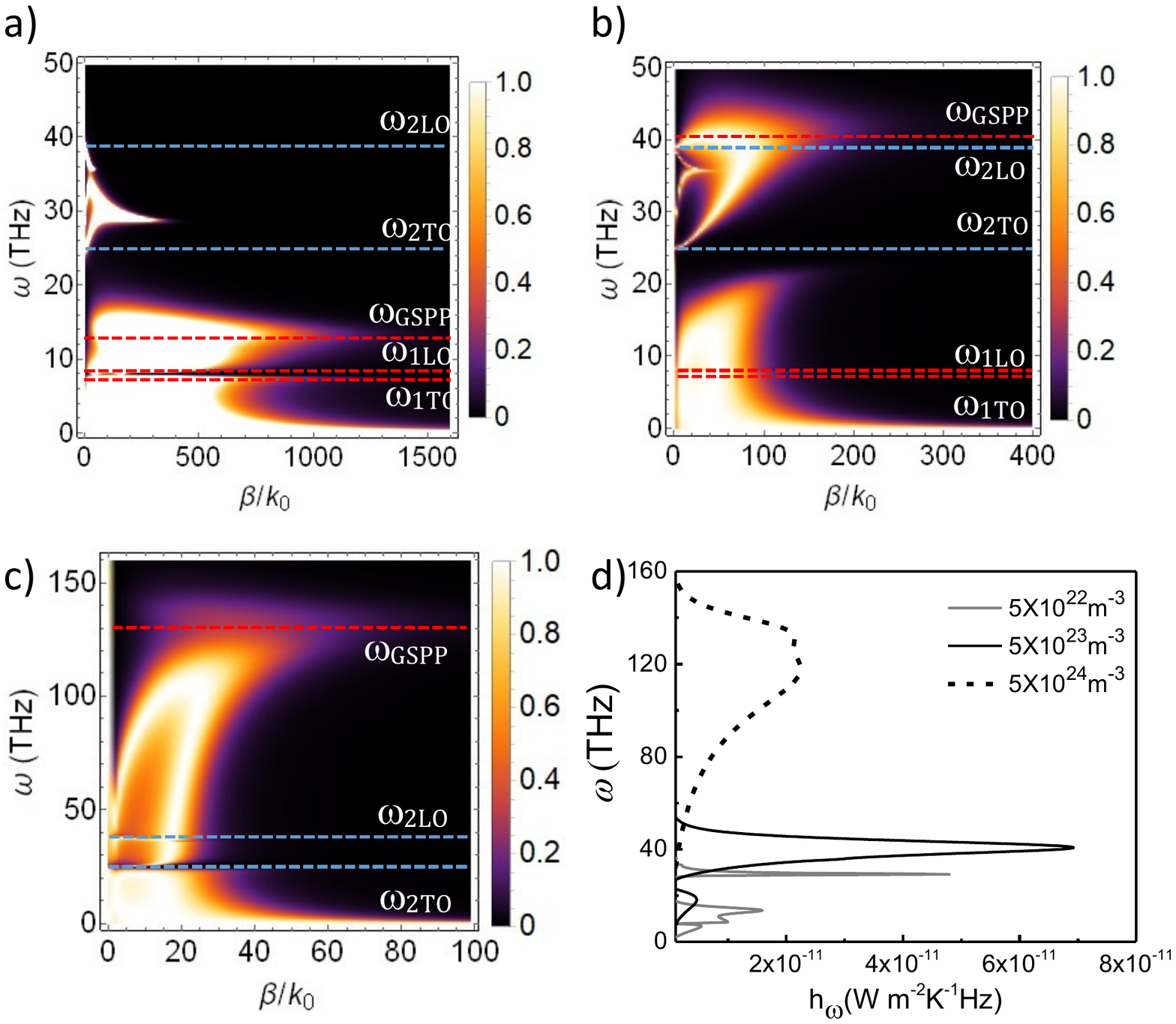}%
\caption{$\tau$ for different values of the carrier concentration $N$.  (a) $N=5\times10^{22}$ m$^{-3}$ ($\omega_p=50.25$ THz), (b) $N=5\times10^{23}$ m$^{-3}$ ($\omega_p=158.92$ THz)and (c) $N=5\times10^{24}$ m$^{-3}$ ($\omega_p=502.54$ THz). (d) Spectral heat transfer coefficient for the different values of $N$.} \label{fig:pull-in}
\end{figure}

To elucidate the interplay between the intragap and interlayer mode couplings in determining the spectral thermal response of the bilayer system we use the quantity usually measured experimentally, i.e., the total heat transfer, $Q$ (Eq. (3)). We define $\eta_{GaAs}=(Q_{GaAs}-Q_{GaAs/NaBr})/Q_{GaAs}$, where $Q_{GaAs}$ is the total heat transfer between two GaAs slabs of thickness $d=50$ $nm$, used as a reference. Figure 5 shows the value of $\eta_{GaAs}$ as a function of the gap separation, $L$ (solid curve). For all $L$ $\eta_{GaAs}$ is positive, so the heat transfer of the bilayer structure is always lower than the reference. We see a peculiar dependence of the total heat transfer with the gap separation $L$. For $L<100$ nm $\eta_{GaAs}$ decreases rapidly towards zero, meaning that the total heat transfer between two bilayers of GaAs/NaBr approaches the one of the GaAs. Thus, the interaction between the two sides of the cavity is dominated by the metallic layer, the GSPP represents the main channel for the total heat transfer and the NaBr plays no crucial role. Plasmons and surface phonons do not couple efficiently in this case and we conclude that for this distance the intragap coupling dominates over the interlayer one. On the other hand, for $L>100$ nm, $\eta_{GaAs}$ reaches a maximum of approximately $30\%$ at a gap separation of $L\sim500$ nm. We also define $\eta_{NaBr}=(Q_{GaAs}-Q_{NaBr/GaAs})/Q_{GaAs}$ that compares the total heat transfer of the bilayer when the order of the layers is reversed, i.e., the top-most layer is NaBr and the backing is GaAs. Again, the reference is $Q_{GaAs}$. In this case, for $L<100$ nm $\eta_{NaBr}\sim1$, meaning that contrarily to the previous case, the contribution from the top-most layer is not important. In this case the interlayer plasmon-phonon coupling dominates over the intragap one. Interestingly, we do not observe a maximum for this layer order with the heat transfer of the bilayer system monotonically approaching the one of GaAs. Both curves go to zero for large $L$ because the systems approach the far-field limit.

The GSPP-hybrid mode competition is explained considering the penetration depth of the surface plasmon polariton at the GaAs-air interface as compared with the thickness $d_1$. The penetration depth is
$\delta=1/Im(\kappa_i)$.  When the gap separation is $L=50$ nm we have $\delta\sim30$ nm, thus $\delta<d_1$. Conversely, when $L=500$ nm, $\delta\sim100$ nm and $\delta>d_1$. The values of the penetration depth depend in general on both frequency and parallel wave vector $\beta$ and are calculated at points where $\tau_p$ is maximum. This analysis allows us to conclude that when the top-most layer is metallic the mode responsible for the heat transfer is very much confined and the system ignores the presence of other backing layers. While, when the top-most layer is dielectric the bilayer is sensitive to the presence of all the layers. 

\begin{figure}[h]
\centering%
\includegraphics[width=0.7\textwidth]{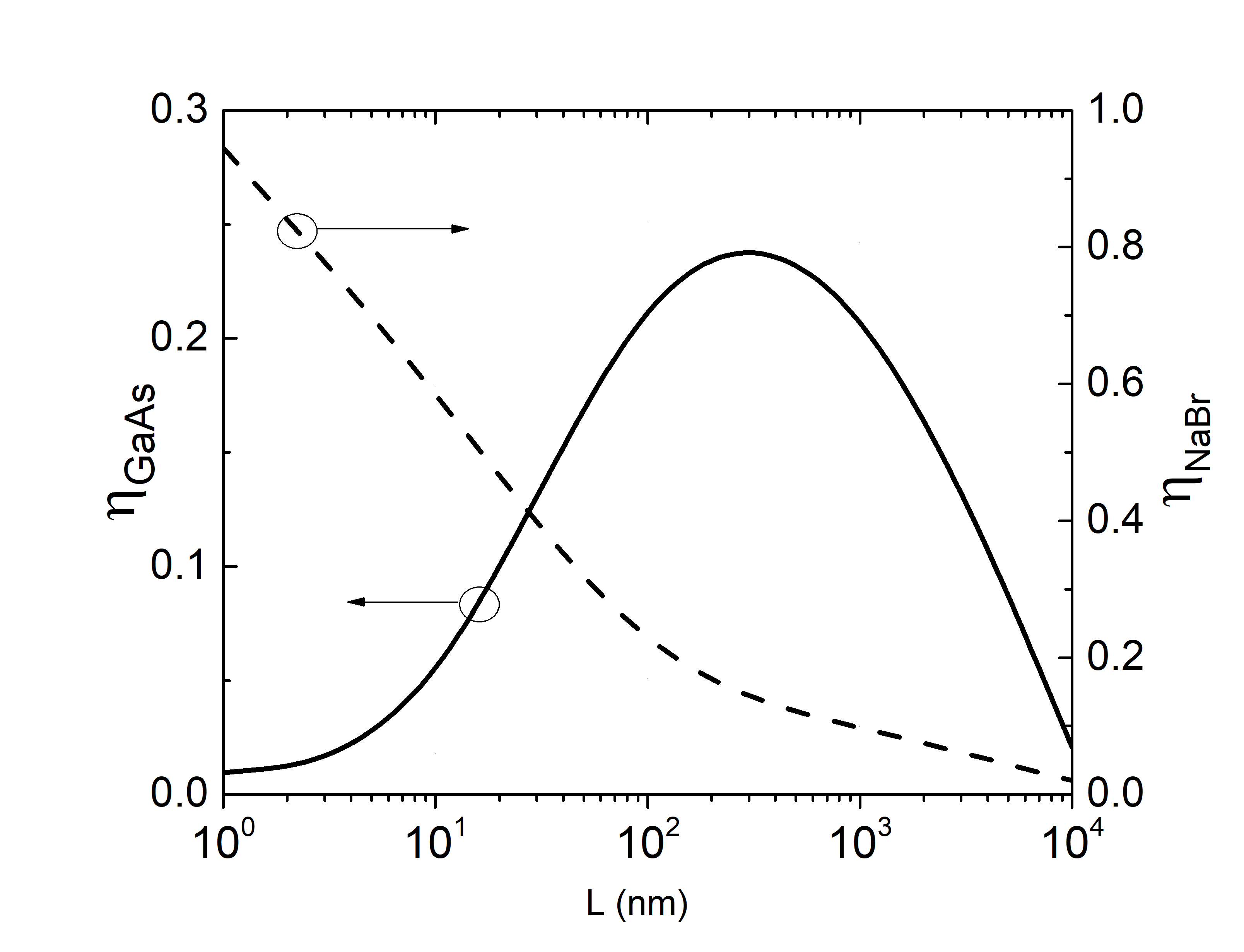}
\caption{Total heat parameters $\eta_{GaAs}=(Q_{GaAs}-Q_{GaAs/NaBr})/Q_{GaAs}$ (solid curve) and $\eta_{NaBr}=(Q_{GaAs}-Q_{NaBr/GaAs})/Q_{GaAs}$ (dashed curve) as a function of the gap separation $L$. The high confinement of the GSPP determines the heat transfer at short separations.} \label{fig:pull-in}
\end{figure}

\section{Conclusions}
The NFRHT can be modified by opening thermal band gaps for specific frequency regions in systems formed by the stacking doped semiconductors and polar dielectric materials. The presence of the gap causes a considerable decrease in the total heat flux for a certain frequency window and all wavevectors. By changing the carrier concentration of the doped semiconductor it is possible to continuously shift the plasmon frequency and thus tune the plasmon-surface phonon coupling. This makes doped semiconductors the ideal material choice to tailor the NFRHT in the infrared efficiently. These gaps are the result of the hybridization of surface plasmon polaritons and surface phonon polaritons and are thus very sensitive to the spatial order of the materials in the system.  Two coupling mechanisms are present in the considered structure: the interlayer coupling at the GaAs/NaBr interface and the intragap one between the two bilayers. The former one is responsible for the opening of the thermal band gap, while the latter one provides the channel for heat transfer across the gap. The latter one dominates over the former one for short gap distances and is characterized by higher confinement of the gap plasmon mode to the point that it becomes insensitive to the surface phonon modes of the polar dielectric. 

 \section{Acknowledgements: }This work was supported by UNAM-DGAPA (PAPIIT IA102117, IN110916), CONACYT-postdoctoral fund 291053, PIIF-UNAM and CIC-UNAM. We thank Philippe Ben-Abdallah for his comments and thorough reading of the manuscript.  \\

G. P. and J. E. P. R. contributed equally to this work.


\end{document}